\documentclass[%
 reprint,
showpacs,
amsmath,amssymb,
 aps,
]{revtex4-1}

\usepackage{fancyhdr}
\usepackage{graphicx}
\usepackage{dcolumn}
\usepackage{bm}

\begin{document}

\title{Plasmon Gun: high-power mid-IR emission at a temporal interface}

\author{Evgenii E. Narimanov }
\affiliation{School of Electrical and Computer Engineering and Birck Nanotechnology Center, Purdue University, West Lafayette, Indiana 47907, USA}
\date{\today}

\begin{abstract}
We propose a mechanism for the generation of intense ultrafast mid-infrared radiation from heavily doped semiconductors. An ultrafast optical pulse transfers a finite impulse to the free carriers, displacing the screening clouds surrounding ionized dopants and inducing coherent plasma-frequency oscillations of the resulting polarization. We derive an exact analytical solution for the corresponding nonequilibrium dynamics and the resulting far-field radiation emitted by a thin semiconductor film. For realistic parameters of heavily doped GaAs, the emitted mid-infrared pulses can reach electric-field amplitudes on the order of $10^7\,{\rm V/m}$ directly in the far field, without relying on optical focusing.
\end{abstract}
\maketitle

\section{Introduction}

The concept of a temporal interface,\cite{TemporalInterface} enabled by advances in ultrafast modulation of material properties,\cite{Single-cycle,MotiVladExpt} has fundamentally expanded the scope of electromagnetic phenomena accessible in time-varying media.\cite{MotiVladReview} By extending temporal modulation rates into the femtosecond domain,\cite{MotiVladExpt} these developments have made it possible to manipulate electromagnetic systems on timescales comparable to their intrinsic dynamical response, giving rise to a broad range of temporally structured media, including photonic time crystals \cite{PT1,PT2,PT3} and related non-stationary electromagnetic systems, and opening access to previously unexplored regimes of wave-matter interaction.\cite{AndreaPT,NaderPT}

Access to intrinsic electronic timescales makes it possible to impart a substantial momentum shift to a free-carrier plasma.\cite{NS1} In a heavily doped semiconductor, however, the free carriers are not merely a conducting background: they also screen the Coulomb potential of ionized impurities.\cite{AMbook} As we show in the present work, a sudden displacement of the electron gas drives the corresponding screening clouds out of equilibrium, resulting in a coherent collective oscillation at the plasma frequency.

The underlying mechanism can be understood by considering the response of a single screened impurity. The sudden displacement of the surrounding screening charge creates a transient dipole moment associated with the impurity. The broad spectral content of this distortion excites the plasma mode of the electron gas, producing a coherent oscillatory response at the plasma frequency. Since the initial momentum shift is common throughout the sample, the resulting plasma oscillation is phase coherent across the excited region, giving rise to a macroscopic polarization and the emission of electromagnetic radiation.

We develop a theoretical description of this process and derive the resulting electromagnetic emission. For realistic parameters of heavily doped GaAs, the plasma frequency naturally falls within the technologically important mid-infrared spectral range,\cite{nmat} and the plasma-frequency oscillation launched by an ultrafast momentum kick produces a coherent electromagnetic pulse at this frequency. We find that the resulting pulse duration is determined by the carrier scattering time, leading to the generation of ultrashort mid-infrared pulses.

An essential advantage of the proposed mechanism is its potentially high emission power, since the carrier densities required to place the plasma frequency in the mid-infrared imply a large density of screened impurities.\cite{nmat} The same conditions that determine the emission frequency therefore ensure a large number of participating emitters, enhancing the collective radiative response.

\section{Doped Semiconductor at a Temporal Interface}

We consider a heavily doped semiconductor layer of thickness d and lateral dimensions much larger than both d and the wavelength corresponding to the plasma frequency, as illustrated in Fig. 1(a). An ultrashort optical pulse is incident normal to the layer, imparting a momentum shift to the free carriers. Throughout this work, we assume that the propagation time of the pump pulse across the semiconductor thickness is small compared to the period of the plasma-frequency oscillation.\cite{MotiVladExpt} Under this condition, the free-carrier momentum shift may be regarded as spatially uniform across the layer, establishing a common initial condition for the carrier dynamics throughout the excited region and leading to the coherence of the subsequent plasma response.

\begin{figure*}[htbp] 
   \centering
   \includegraphics[width=6.5 in]{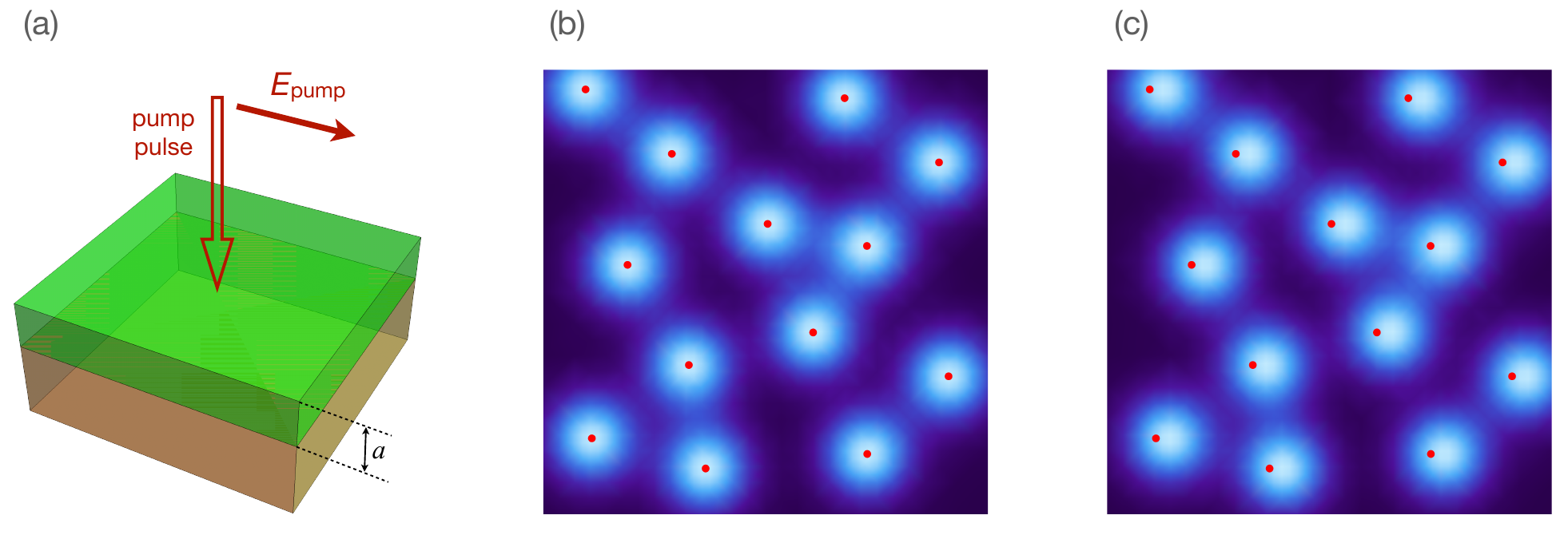} 
   \caption{Physical picture of the proposed mechanism. (a) A heavily doped semiconductor film is driven by an ultrafast optical pulse incident normal to the film. The film thickness $a$ is smaller than the wavelength corresponding to the plasma frequency, ensuring a phase-coherent carrier response across the layer. (b) Equilibrium screening clouds (with free carrier density shown in false color) surrounding the ionized dopants (red dots). (c) Immediately after the pump pulse, the free carriers acquire a finite momentum, displacing the screening clouds relative to the stationary dopants and creating transient dipole moments associated with the screened impurities.}
   \label{fig:1}
\end{figure*}

In equilibrium, the Coulomb potential of each ionized donor is screened by the surrounding electron gas,\cite{AMbook} producing a localized screening cloud characterized by the Thomas-Fermi screening length $r_s$, as illustrated in Fig. 1(b). For the doping densities considered here, the screening clouds associated with neighboring impurities remain largely distinct, allowing each donor to be viewed as being surrounded by its own localized screening charge distribution.\cite{ZimanEP}

Immediately following the ultrafast excitation, the ionized donor lattice remains stationary while the free carriers acquire a finite momentum.\cite{AMbook} The corresponding screening charge distributions are therefore displaced relative to their equilibrium positions, producing the distorted screening clouds illustrated in Fig. 1(c). As we show below, this sudden distortion creates a transient dipole moment associated with each screened impurity and initiates the subsequent dynamics of the free-carrier plasma.

In equilibrium, the ionized dopants are screened by the surrounding free carriers.\cite{ZimanEP} The corresponding screening charge distribution can be described in terms of the effective single-particle potential \cite{AMbook}
\begin{equation}
V_{\rm eff}({\bf r} )
=
\frac{4\pi r_s^2}{\epsilon_\infty} n_0 e
-
\frac{e}{\epsilon_\infty}
\sum_n
\frac{
e^{- \left|{\bf r} -{\bf r} _n \right|/r_s}
}{
|{\bf r} - {\bf r} _n|
},
\label{eq:Veff}
\end{equation}
where $n_0$ is the free-carrier density and
\begin{equation}
r_s
=
\sqrt{
\frac{\epsilon_\infty \varepsilon_F}
{6\pi n_0 e^2}
},
\label{eq:rs}
\end{equation}
is the screening radius.

In the proposed approach, the heavily doped semiconductor is driven by an ultrafast optical pulse. Such excitation makes it possible to employ pulses containing only a small number of oscillations of the electric field, which can transfer a finite momentum ${\bf p}_0$
to the free carriers.\cite{NS1} In high-mobility materials such as GaAs, the period of the optical field is much shorter than the carrier scattering time.\cite{nmat} As a result, when the pulse contains only a few optical cycles,\cite{MotiVladExpt} the momentum transfer occurs on a timescale over which collisions may be neglected \cite{nmat} and therefore appears as an abrupt displacement of the carrier momentum distribution.\cite{AMbook} The carrier distribution immediately after the pulse at $t=0$ is then given by
\begin{equation}
f\left({\bf p},{\bf r}; t = 0\right)
=
f_0
\left(
\varepsilon_{{\bf p}-{\bf p}_0}
+
e V_{\rm eff}({\bf r} )
\right)
\label{eq:f}
\end{equation}
Then for times $t$ shorter than the scattering time $\tau$,
\begin{equation}
0 <  t \ll \tau,
\label{eq:t}
\end{equation}
the carrier distribution function satisfies the collisionless kinetic equation \cite{ZimanEP}
\begin{equation}
\frac{\partial f}{\partial t}
+
{\bf v}_{{\bf p}}\cdot\nabla f
+
e{\bf E}({\bf r}, t)\cdot
\frac{\partial f}{\partial {\bf p}}
=
0
\label{eq:KE}
\end{equation}
where the electric field ${\bf E}$ is defined self-consistently by  the Poisson equation \cite{AMbook}
\begin{eqnarray}
\nabla\cdot{\bf E}({\bf r} ,t)
&=&
-\frac{4\pi e}{\epsilon_\infty}
\sum_n \delta({\bf r} -{\bf r} _n) 
\nonumber \\
& + & 
\frac{4\pi e}{\epsilon_\infty}
\frac{2}{(2\pi\hbar)^3}
\int d^3{\bf p}\,
f({\bf p},{\bf r} ;t).
\label{eq:Poisson}
\end{eqnarray}

\begin{figure*}[htbp] 
   \centering
   \includegraphics[width=7 in]{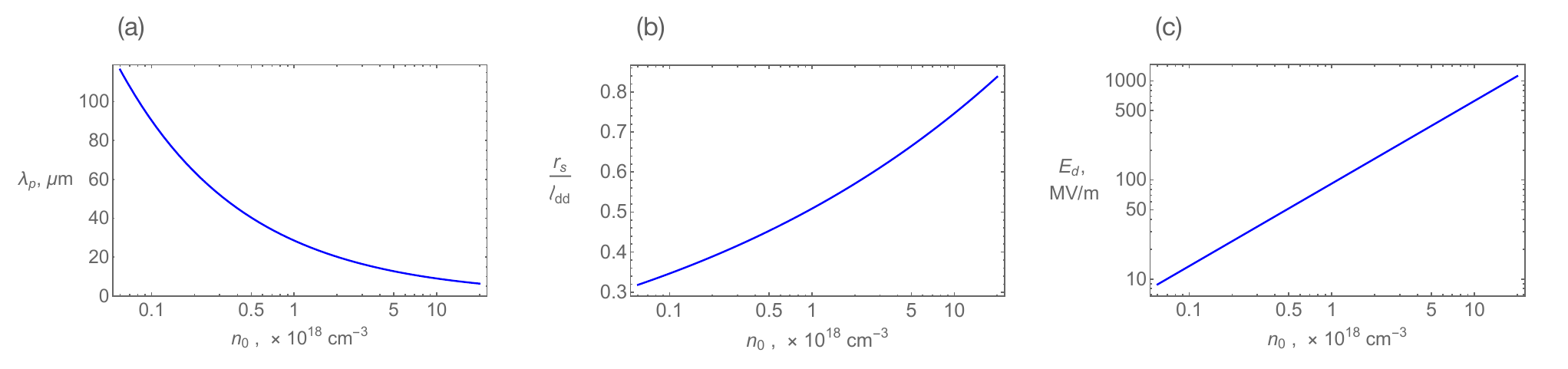} 
   \caption{Material parameters governing the proposed mechanism in GaAs as functions of the carrier concentration $n_0$. (a) Plasma wavelength $\lambda_p$. (b) Ratio $r_s/\ell_{dd}$ of the screening radius to the average dopant-dopant separation. The condition $r_s < \ell_{dd}$ is essential for the formation  of well-defined screening clouds around individual dopants. It therefore determines the range of validity of the present theory and imposes a lower limit on the achievable emission wavelength. (c) Characteristic doping field $E_d$, which sets the natural field scale of the emitted radiation.}
   \label{fig:2}
\end{figure*}

Substituting the solution of the linear kinetic equation (\ref{eq:KE}) into
the Poisson equation (\ref{eq:Poisson}) and 
introducing the scalar potential $\varphi$, we obtain
\begin{eqnarray}
\varphi_{{\bf k}}(t)
& + &
\frac{1}{(kr_s)^2}
\int_0^t dt' \,
\varphi_{{\bf k}}(t') \nonumber \\
& \times & 
e^{i \frac{{\bf k}\cdot{\bf p}_0}{m_0}\left(t' - t\right)}
\frac{d}{dt'}\, {\rm sinc}(kv_F\left(t'-t\right)) \nonumber \\
& = &
- \frac{e\, \sum_n e^{-i{\bf k}\cdot{\bf r}_n}}{ 2\pi^2 \epsilon_\infty k^2}
- \frac{V_{\bf k}\, e^{-i \frac{{\bf k}\cdot{\bf p}_0}{m_0}t}}{k^2 r_s^2}
{\rm sinc}(kv_Ft).
\label{eq:intEqn}
\end{eqnarray}
where
\begin{eqnarray}
V_{\bf k}
& = &
- \frac{4\pi e}{\epsilon_\infty}
\frac{\sum_n e^{-i {\bf k}\cdot{\bf r}_n}}{k^2+1/r_s^2},
\label{eq:Vk}
\end{eqnarray}
is the Fourier transform of $V_{\rm eff}\left({\bf r}\right)$.

Introducing the transformation
\begin{equation}
\varphi_{{\bf k}}(t) = \left(  V_{\rm eff}({\bf k}) + \psi_{{\bf k}}(t) \right) \, 
e^{-i \frac{{\bf k}\cdot{\bf p}_0}{m_0} t},
\label{eq:varphi}
\end{equation}
we reduce (\ref{eq:intEqn}) to the Volterra integral equation
\begin{eqnarray}
\psi_{{\bf k}}(t)
& + &
\frac{1}{(kr_s)^2}
\int_0^t dt' \,
\psi_{{\bf k}}(t')
\frac{d}{dt'}{\rm sinc}\left(k v_F \left(t' - t\right)\right)
\nonumber \\
&=&
- \frac{e}{2\pi^2\epsilon_\infty k^2}
\sum_n e^{-i{\bf k}\bar{r}_n}
\left(e^{i \frac{{\bf k}\cdot{\bf p}_0}{m_0} t}-1\right), \ \ \ \ 
\label{eq:psiEqn}
\end{eqnarray}
which allows an exact analytical solution.

Despite the availability of an exact analytical solution, the physically relevant regime corresponds to a momentum transfer that remains small compared to the Fermi momentum, $p_0 \ll p_F$. Indeed, even for strong optical excitation, the oscillatory character of the electric field leads to substantial cancellation of the transferred momentum,\cite{BonWolf} making a large net momentum shift difficult to achieve. We therefore restrict our analysis to the leading-order response in $p_0/p_F$, which captures the essential physics of the effect while substantially simplifying the resulting expressions.

In the leading order in, ${p}_0/p_F$, we obtain
\begin{equation}
\psi_{{\bf k}}(t)
=
- \frac{ie}{2\pi^2\epsilon_\infty}
\, 
\frac{({\bf k}\cdot{\bf p}_0)}{p_Fk^3}
\, 
\sum_n e^{-i{\bf k}\bar{r}_n}
\, 
\phi_k\left(k v_F t \right),
\label{eq:psiSol}
\end{equation}
where the dimensionless function $\phi_k\left(u\right)$ satisfies the equation
\begin{equation}
\phi\left(u\right)+ \frac{1}{(kr_s)^2}\int_0^u d\xi\,\phi(\xi)\,
{\rm sinc}'\left(\xi - u\right) = u,
\label{eq:phiEqn}
\end{equation}
with exact  solution
\begin{eqnarray}
\phi_k\left(u\right)
&=&
 \int_0^1 \frac{du}{u}\,
\frac{(kr_s)^2  \, \sin(ux)}{\left( (kr_s)^2+1+u\, {\rm arctanh}\,u \right)^2 +(\pi u/2)^2}
\nonumber\\
& + & \frac{(kr_s)^2}{1+ (kr_s)^2}  \, u + 
\frac{2 \,  (kr_s)^2 }{ \beta_k^2 \, C_{k}} \sin(\beta_k u)
\label{eq:46}
\end{eqnarray}
where
\begin{equation}
C_{k} = \frac{\beta_k}{\beta_k^2-1} - 
{\rm arctanh}\, \frac{1}{\beta_k},
\label{eq:47}
\end{equation}
and $\beta_k$ is defined by
\begin{equation}
\beta_k \, {\rm arctanh}\,\frac{1}{\beta_k} = 1+(kr_s)^2.
\label{eq:48}
\end{equation}
At distances larger than the screening radius from a dopant, $\left| {\bf r} - {\bf r}_n\right| \gg r_s$, the resulting scalar potential can be expressed in terms of the field of a point dipole, 
\begin{eqnarray}
\varphi(\bar{r},t)
& = &
\frac{\left({\bm  \mu}_0\cdot{\bm r}\right)}{\epsilon_\infty r^3}
\sin(\omega_p t),
\label{eq:phiSol}
\end{eqnarray}
where the dipole moment
\begin{equation}
{\bm \mu}_0\equiv \frac{{\bm p}_0}{ p_F} e r_s .
\label{eq:mu}
\end{equation}

The polarization of the entire sample is therefore given by
\begin{equation}
{\bm P} = n_0\,{\bm \mu}_0\, \sin\left(\omega_p t\right).
\label{eq:P}
\end{equation}

\section{Plasma-Frequency Radiation}

With the ultrashort pump pulse inducing coherent oscillations of the free-carrier polarization at the plasma frequency [see Eqn. (\ref{eq:P})], a thin film of heavily doped semiconductor acts as a source of the corresponding electromagnetic radiation emitted into the far field. The thin-film requirement ensures that the propagation delay across the semiconductor layer remains much shorter than the period of the emitted radiation. As a result, the polarization oscillations throughout the film remain nearly in phase, allowing the entire layer to radiate coherently as a single source.

We consider a semiconductor slab of thickness $a$ smaller than the plasma wavelength $\lambda_p$, occupying the region $0 < z < a < \lambda_p$, on top of a transparent substrate with dielectric permittivity $\epsilon_s$. The modulation pulse is incident along the $z$-direction and transfers a momentum ${\bf p}_0$ to the free carriers in the plane of the film. The resulting free-carrier polarization is therefore uniform throughout the layer and oriented parallel to ${\bf p}_0$ -- see Eqn. (\ref{eq:P}).

A straightforward solution of Maxwell's equations in this geometry yields the far-field electric-field amplitude of the radiation emitted at the plasma frequency,
\begin{equation}
E
=
-\frac{4\pi}{c}
\frac{\mu_0 n_0 \omega_p a}
     {1+\sqrt{\epsilon_s}}
\label{eq:E}
\end{equation}
or, equivalently in SI units, 
\begin{eqnarray}
E
&=&
-\frac{2\pi}{\epsilon_0}
\frac{a}{\lambda_p}
\frac{p_0}{p_F}
\frac{e r_s n_0}
     {1+\sqrt{\epsilon_s}},
\label{eq:ESI}
\end{eqnarray}

The magnitude of the emitted field is therefore set by the semiconductor doping scale
\begin{equation}
E_d \equiv \frac{e r_s n_0}{\epsilon_0} \sim 10^{8}\ {\rm V/m},
\end{equation}
reduced by the ratio $p_0/p_F$
 of the momentum impulse transferred by the pump to the Fermi momentum.

\section{Practical Implementation and Tunability}

As a representative implementation of the proposed approach, we consider a heavily doped GaAs film. Owing to its high carrier mobility, well-established growth technology, and the availability of high-quality material over a broad range of doping levels,\cite{nmat} GaAs provides a convenient platform for the generation of plasma-frequency radiation. Furthermore, for carrier concentrations in the range $10^{18} - 10^{19}\,{\rm cm}^{-3}$, the corresponding plasma wavelengths fall within the mid-infrared spectral range,\cite{nmat} where ultrafast optical excitation and radiation detection are both readily accessible.

The key material parameters governing the proposed mechanism are the plasma wavelength $\lambda_p$, the ratio $r_s/\ell_{dd}$ of the screening radius \cite{AMbook} to the average dopant-dopant separation, and the characteristic doping field $E_d \equiv e r_s n_0/\epsilon_0$. Fig. 2 shows the dependence of these quantities on the carrier concentration $n_0$ for GaAs. Increasing the doping level decreases the plasma wavelength, moving the emitted radiation deeper into the mid-infrared spectral range, while simultaneously increasing the ratio $r_s/\ell_{dd}$. As the proposed mechanism relies on the existence of well-defined screening clouds surrounding individual dopants, its applicability is restricted to the regime $r_s < \ell_{dd}$. This condition therefore imposes a lower limit on the achievable emission wavelength. At the same time, the characteristic field scale $E_d$ increases rapidly with carrier density, reaching values on the order of $10^8\,{\rm V/m}$ for $n_0\sim10^{18}\,{\rm cm}^{-3}$.

The remaining factor controlling the emitted field amplitude is the ratio $p_0/p_F$ of the impulse transferred by the pump pulse to the Fermi momentum. Unlike the characteristic field scale $E_d$, which is determined by the material parameters, $p_0/p_F$ depends on the excitation conditions. Because the optical electric field changes sign during the pulse, the contributions of different optical cycles partially cancel, making a large net momentum shift difficult to achieve even at high pump intensities. Nevertheless, few-cycle excitation can produce a finite impulse transfer, and values of $p_0/p_F$ in the range $10^{-2}$ to $10^{-1}$ appear realistic without requiring extreme excitation conditions.

For a representative GaAs thin film ($a = 5 \, \mu{\rm m}$) with $n_0\sim10^{18}\,{\rm cm}^{-3}$, the characteristic doping field is $E_d\sim10^8\,{\rm V/m}$, while the corresponding plasma wavelength is of the order of several tens of microns. Taking a conservative value $p_0/p_F\sim 0.1$ and the film thickness $a \lesssim \lambda_p/3$, Eq. (\ref{eq:ESI}) yields an emitted field amplitude on the order of $10^7\,{\rm V/m}$. 

Remarkably, this value corresponds to the amplitude of the emitted far-field radiation itself rather than to a focused pump field. Such field strengths are therefore attainable directly in a freely propagating mid-infrared pulse, highlighting the potential of the proposed mechanism as a compact source of intense coherent radiation at the plasma frequency.

\begin{figure}[htbp] 
   \centering
    \includegraphics[width=3in]{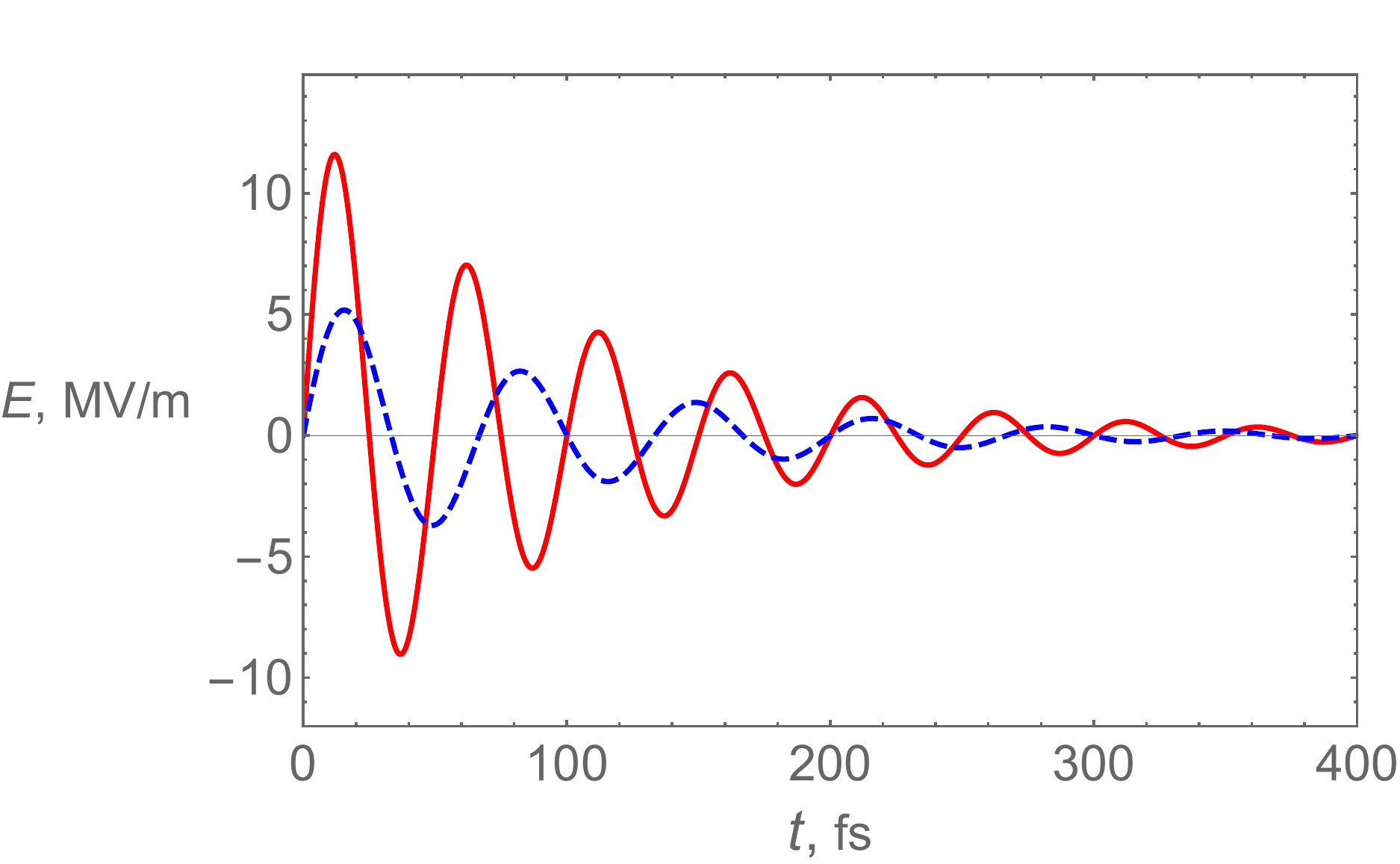} 
   \caption{Temporal profile of the mid-infrared radiation emitted from an $n$-doped GaAs thin film with plasma wavelengths  $\lambda_p=15\,\mu{\rm m}$ (red solid line) and $\lambda_p=20\,\mu{\rm m}$ (blue dashed line), corresponding to carrier densities $n_0 = 4.6\times10^{18}\,{\rm cm}^{-3}$ and $n_0=2.6\times10^{18}\,{\rm cm}^{-3})$ respectively. The film thickness is $a=5\,\mu{\rm m}$, and the substrate is semi-insulating GaAs ($\epsilon_s=12.9$). The pulse envelope is determined by the free-carrier scattering time, while the carrier frequency is set by the plasma frequency of the doped semiconductor.}
   \label{fig:3}
\end{figure}

The temporal profile of the emitted radiation is illustrated in Fig. 3 for two representative plasma wavelengths, $\lambda_p=15\,\mu{\rm m}$ and $\lambda_p=20\,\mu{\rm m}$. In both cases, the emission consists of a short mid-infrared pulse containing only a few oscillations of the electric field. The pulse duration is determined by the free-carrier scattering time, which limits the lifetime of the coherent plasma oscillations. Nevertheless, for realistic scattering times in high-quality GaAs,\cite{nmat} the emitted waveform retains a well-defined carrier frequency while remaining confined to a sub-picosecond time window.

\section{Conclusions}

We have shown that an ultrafast optical pulse can induce coherent plasma-frequency radiation from a heavily doped semiconductor through the transient displacement of the screening clouds surrounding ionized dopants. The resulting non-equilibrium state is characterized by a finite impulse transferred to the free carriers, which generates oscillating dipole moments associated with the screened impurities and drives coherent polarization oscillations at the plasma frequency.

For a thin semiconductor film, these oscillations act as a source of coherent electromagnetic radiation emitted into the far field. For realistic parameters of heavily doped GaAs, the resulting coherent mid-infrared radiation can reach far-field amplitudes on the order of $10^7\,{\rm V/m}$, despite being emitted directly into free space rather than generated by focusing an external optical field.

These results identify a new mechanism for the generation of intense ultrafast mid-infrared radiation and suggest heavily doped semiconductor films as a promising platform for compact ultrafast mid-infrared emitters.

\section{Acknowledgements}

The author thanks Boris Shapiro for discussions that helped motivate this work and for valuable comments on the manuscript.

\end{document}